\documentstyle[12pt]{article}
\begin{document}
\baselineskip=22pt plus 0.2pt minus 0.2pt
\lineskip=22pt plus 0.2pt minus 0.2pt
\font\bigbf=cmbx10 scaled\magstep3
\begin{center}
 {\bigbf Fock representations from $U(1)$ holonomy algebras}\\

\vspace*{0.35in}

\large

Madhavan Varadarajan
\vspace*{0.25in}

\normalsize

{\sl Raman Research Institute,
Bangalore 560 080, India.}
\\
madhavan@rri.ernet.in\\
\vspace{.5in}
November 1999\\
\vspace{.5in}
ABSTRACT

We revisit the quantization of $U(1)$ holonomy algebras using the abelian
$C^*$ algebra based techniques which form the mathematical underpinnings 
of current efforts to construct loop quantum gravity. In particular,
we clarify 
the role of ``smeared loops'' and of Poincare invariance in the construction 
of  Fock 
representations of these algebras. This enables us to critically re-examine 
early pioneering efforts to construct Fock space representations of
 linearised gravity and free Maxwell theory from holonomy algebras through 
an application 
of the (then current) techniques of loop quantum gravity.

\end{center}

\pagebreak

\setcounter{page}{1}

\section*{1. Introduction} 

In the early nineties \cite{lingrav, carlomax, extloop} linearised gravity 
in terms of connection variables and 
 free Maxwell theory on flat spacetime, were treated as 
useful toy models on which to test techniques being developed for loop
quantum gravity\cite{carlolee} . 
Significant progress has been made in the field of loop
quantum gravity since then\cite{almmt}. Hence, 
it is useful to 
revisit these systems using current techniques to clarify  certain 
questions which arise in the context of  those pioneering but 
necessarily non-rigorous efforts.

Two important (and related) questions are:

\noindent (I) How did  similar techniques  for the quantization of 
general relativity  and  for its linearization about flat space,
result in a non Fock representation for the (kinematic sector) of the 
former and a Fock representation
for the latter? In particular, what is the role of Poincare 
invariance in obtaining the Fock representation? (This last point was a 
puzzle to the authors themselves \cite{lingrav}).

\noindent (II)What is the role of ``smeared'' loops in \cite{lingrav}
in obtaining a Fock 
representation?

In this work, we use the abelian $C^*$ algebra techniques \cite{ai, al}
which constitute the mathematically rigorous 
 framework of the loop quantum gravity program today, to investigate 
(I) and (II) above. It is also our aim to clarify the role of the different
mathematical structures in the quantization procedure 
which determine whether a  Fock or non Fock  
  representation results.
Although we restrict attention to $U(1)$ theory
on a flat spacetime, we believe that our results should
be of some relevance  to the 
case of linearised gravity.

This work is motivated by the following question in loop quantum gravity: how
do Fock space gravitons on flat spacetime arise from the non-Fock structure 
of the Hilbert space which serves as the kinematical arena for loop 
quantum gravity? Admittedely, the answer to this question must await the 
construction of the full physical state space (i.e. the kernel of all the 
constraints) of quantum gravity. Nevertheless, this work may illuminate
some facets of the issues involved.

The starting point for our analysis is the 
abelian Poisson bracket 
algebra of $U(1)$ holonomies around loops on a spatial slice.
This algebra is completed to the abelian $C^*$ algebra,
$\overline{\cal{HA}}$ 
of \cite{ai,al}. Hilbert space representations of 
$\overline{\cal{HA}}$  are in 
determined by continuous 
positive linear functions (PLFs) on $\overline{\cal{HA}}$. We review the 
construction of $\overline{\cal{HA}}$  and of the  PLF introduced 
in \cite{ai,al}  (which we shall call the Haar PLF) in section 2.
The resulting representation is a non Fock representation in which the 
Electric flux is quantized \cite{aimax}. 

In section 3 we construct an abelian $C^*$ algebra 
${\overline{\cal{HA}}}_r$, based on the Poisson bracket algebra of holonomies 
around the ``Gaussian smeared '' loops of \cite{lingrav}.\footnote{ $r$
is a small length which characterises the width of the Gaussian smearing 
function in \cite{lingrav}.}
Next, we derive the key result of this work, namely that there exists a 
natural $C^*$ algebraic isomorphism, 
$I_r:{\overline{\cal{HA}}} \rightarrow {\overline{\cal{HA}}}_r$ with the 
property that $I_r({\cal{HA}}) ={\cal{HA}}_r$.

The standard flat spacetime Fock vacuum expectation value restricts to a
positive linear function on ${\cal{HA}}_r$. We are unable to show the 
continuity or lack thereof, of this Fock PLF on ${\cal{HA}}_r$.
Nevertheless, since the GNS construction needs only a * algebra (as opposed
to a $C^*$ algebra), we can use the Fock PLF to construct a representation of 
 the * algebra ${\cal{HA}}_r$. In section 4 we show that this representation
is indeed the standard Fock representation even though  ${\cal{HA}}_r$
is a proper subalgebra of the standard Weyl algebra for $U(1)$ theory.

Using the map, $I_r$, we can define a Haar PLF on 
${\overline{\cal{HA}}}_r$. 
We construct the resulting representation in section 5a.
Finally, we use $I_r$ to define a Fock PLF on ${\cal{HA}}$ . The 
resulting representation is, in a precise sense, an approximation to the 
standard Fock representation. We study it in section 5b.

Section 6 is devoted to a discussion of our results in the context of the
questions (I) and (II). Some useful lemmas are proved in Appendices A1 and
A2.

In  this work the spacetime of interest is flat $R^4$ and we  use
 global cartesian coordinates $(t,x^i), \; i=1,2,3$. The spatial slice of 
interest is  the initial $t=0$ slice and
all calculations are done 
in the spatial cartesian coordinate chart $(x^i)$. 
We use units in which both 
the velocity of light and Plancks constant, $\hbar$,
are equal to 1.
We  freely raise and lower indices with the flat spatial metric.
The Poisson bracket 
between the $U(1)$ connection $A_a({\vec x}), a= 1,2,3$ and its  
conjugate electric field $E^b({\vec y})$ is 
$\{A_a({\vec x}),E^b({\vec y})\} =e\delta_a^b\delta({\vec x},{\vec y})$
where $e$ is a constant with units of electric charge.

\section*{2. Review of the construction and representation theory
of $\overline{\cal{HA}}$.
         }
We quickly review the relevant contents of \cite{ai,al}. 
We refer the reader to \cite{ai,al}, especially appendix A2 of 
\cite{al} for details.

The mathematical structures of interest are as follows.

$\cal{A}$ is the space of smooth $U(1)$ connections on the trivial 
$U(1)$ bundle on $R^3$.\footnote {Thus a minor change of notation from 
A2 of \cite{al} is that we denote ${\cal{A}}_0$ of that reference by
by $\cal{A}$.} We restrict attention to connections $A_a(x)$
whose cartesian components are functions of rapid decrease at infinity.

${\cal L}_{x_0}$ is the space of unparametrized, oriented, piecwise 
analytic loops  
\footnote{This is in contrast to the $C^1$ loops of A2 of \cite{al}.}
on $R^3$ with basepoint $\vec{x_0}$. 
Composition of a loop $\alpha$ with a loop $\beta$ is denoted by 
$\alpha\circ \beta$.
Given a loop $\alpha \in {\cal L}_{x_0}$,
the holonomy of $A_a(x)$ around $\alpha$ is 
$H_\alpha (A):= \exp (i \oint_{\alpha} A_a dx^a)$.

$\tilde{\alpha}$ is the holonomy equivalence class (hoop class) of 
$\alpha$ i.e. $\alpha,\beta$ define the same hoop iff 
$H_\alpha (A) =H_\beta (A)$
 for every $A_a(x)\in {\cal A} $.

$\cal{HG}$ is the group generated by all hoops $\tilde{\alpha}$, where
group multiplication is hoop composition i.e. 
$\tilde{\alpha}\circ \tilde{\beta} := \widetilde{\alpha\circ \beta}$. 

${\cal{HA}}$ 
is the abelian Poisson bracket algebra of $U(1)$ 
holonomies.

${\cal FL}_{x_0}$ is the free algebra generated by elements of  
${\cal L}_{x_0}$, with product law $\alpha \beta :=\alpha\circ \beta$.
With this product, all elements of 
${\cal FL}_{x_0}$ are expressible as complex linear combinations of 
elements of ${\cal L}_{x_0}$. 

$K$ is a 2 sided ideal of ${\cal FL}_{x_0}$, such that 
\begin{equation}
\sum_{i=1}^N a_i \alpha_i \in K \;{\rm iff}\;\;
\sum_{i=1}^N a_i H_{\alpha_i}(A) = 0\;  {\rm for\; every} \;
A_a(x) \in {\cal A},
\end{equation}
where $a_i$ are complex numbers. 

${\cal FL}_{x_0}$ is quotiented by $K$ to give the algebra 
${\cal FL}_{x_0}/ K$. 
The $K$ equivalence class of $\alpha$ is denoted by 
$[\alpha ]$.
As abstract algebras, ${\cal{HA}}$  and ${\cal FL}_{x_0}/K $
are isomorphic. 
\begin{equation}
(\sum_{i=1}^N a_i [\alpha_i ])^* := \sum_{i=1}^N a_i^*[ \alpha_i^{-1}] 
\end{equation}
defines a $*$ relation on ${\cal{HA}}$.
\begin{equation}
||\sum_{i=1}^N a_i [\alpha_i ]|| := 
\sup_{A\in {\cal A}} |\sum_{i=1}^N a_i H_{\alpha_i}(A)|
\end{equation}
defines a norm on ${\cal{HA}}$. $\overline{\cal{HA}}$ is the abelian 
$C^*$ algebra obtained by defining $*$ on ${\cal{HA}}$ and completing 
the resulting $*$ algebra with respect to $||\;\;\;||$.

$\Delta$ is the spectrum of $\overline{\cal{HA}}$.
$\Delta$ is also denoted by 
$\overline{\cal A/G}$ where $\cal G$ denotes the $U(1)$ gauge group and is 
a suitable completion of the space of connections modulo gauge, 
${\cal A/G}$. From Gel'fand theory, 
$\Delta$ is the space of continuous, linear, multiplicative $*$
homeomorphisms, $h$, from $\overline{\cal{HA}}$ to the ($C^*$ algebra of)
 complex numbers $\bf C$. From \cite{al} the elements of 
$\Delta$ are also in 1-1 correspondence with homeomorphisms from
${\cal HG}$ to $U(1)$. 

Given $X\in \overline{\cal{HA}} $, 
$h(X)$ is a complex function on 
$\Delta$. $\Delta$ is endowed with the weakest topology 
in which $h(X)$ for all $X\in \overline{\cal{HA}} $  
are continuous functions on $\Delta$. In this topology,
$\Delta$ is a compact, Hausdorff space and the
functions $h([\alpha]),\;
\alpha \in {\cal L}_{x_0}$ are dense in the $C^*$ algebra, $C(\Delta )$, 
of continuous
functions  on $\Delta$. Further,
$C(\Delta )$ is isomorphic to $\overline{\cal{HA}}$. 
Every continuous cyclic representation of 
$\overline{\cal{HA}}$ is in 1-1 correspondence with a continuous 
positive linear functional (PLF) on $\overline{\cal{HA}}$.
Since $\overline{\cal{HA}} \cong C(\Delta )$, every continuous PLF so 
defined on $C(\Delta )$ is in correspondence, by the Riesz lemma, with 
some regular measure $d\mu$ on $\Delta$ and ${\hat {H}}_{\alpha}$
is represented on $\psi \in L^2(\Delta, d\mu)$ as unitary operator through
$({\hat {H}}_{\alpha}\psi)(h) = h([\alpha ]) \psi (h)$.

In particular, the continuous `Haar' PLF \cite{al}
\begin{eqnarray}
\Gamma(\alpha )& =& 1\; {\rm if}\; \tilde{\alpha} =\tilde{o}\nonumber \\
               & = & 0\;\;\;{\rm otherwise}
\label{eq:haarplf}
\end{eqnarray}
(where $o$ is the trivial loop), corresponds to the Haar measure
on $\Delta$. 

 $\Delta =\overline{\cal A/G}$ can also be 
constructed as the  projective limit space \cite{mm}
of certain finite dimensional 
spaces. Each of these spaces 
is isomorphic to $n$ copies of $U(1)$
and is labelled by $n$ strongly independent hoops. 
Recall from \cite{al} that $\tilde{\alpha}_i 
\;i=1..n$  are strongly independent 
hoops iff $\alpha_i \in {\cal L}_{x_0}$ are strongly independent loops;
$\alpha_i , \; i=1..n$ are strongly independent loops iff each $\alpha_i$ 
has at least one segment which intersects $\alpha_{j\neq i}$ at most at 
a finite number of points.
The Haar measure on $\Delta$ is the projective limit measure of the 
Haar measures  on each of the finite dimensional spaces.
\footnote{ Note that the proof of continuity of the Haar PLF in \cite{al}
is incomplete in that it applies only if the loops $\alpha_j$ of A.7 of 
\cite{al} are holonomically independent. Nevertheless, if as in this work,
we restrict attention to piecewise analytic loops, continuity of
the Haar PLF  can 
immediately be inferred from its definition through  the Haar measure.}
Then the 
considerations of \cite{alproj} show that the electric flux 
$\int_S E^a ds_a$
through a surface $S$ can be realised as an essentially self adjoint 
operator
on the dense domain of cylindrical functions 
\footnote{Cylindrical functions on $\Delta$ are of the form 
$\psi_{\{[\alpha_i]\}}:=\psi(h([\alpha_1])..h([\alpha_n]))$,
 where $\alpha_i, i=1..n$, are a finite
number of strongly indendent loops and $\psi$ is a complex function on 
$U(1)^n$.}
as
\begin{equation}
\int_S {\hat E}^a ds_a \psi_{\{[\alpha_i]\}} = e\sum_{i}N(S,\alpha_i ) 
  h([\alpha_i]){\partial \psi_{\{[\alpha_i]\}}\over \partial h([\alpha_i])}  
\end{equation}
where $N(S,\alpha_i )$ 
is the number of intersections between $\alpha_i$ and $S$.

\section*{3. ${\overline{\cal{HA}}}_r$ and the isomorphism $I_r$}

In section 3a we recall the definition of `smeared' loops and their 
holonomies from \cite{lingrav} and construct the `smeared' loop related 
structures $\tilde{\alpha}_r$, $K_r$, ${\cal{HA}}_r$, $\overline{\cal{HA}}_r$
and $\Delta_r$. In section 3b, using Appendix A2, we 
show that an isomorphism exists between the structures
$\tilde{\alpha}$, $K$, ${\cal{HA}}$, $\overline{\cal{HA}}, \Delta$ 
and their 
`smeared' versions.

\subsection*{3a. The construction of ${\overline{\cal{HA}}}_r$}
In the notation of \cite{lingrav},
\begin{eqnarray}
H_{\alpha}(A) & = & \exp i\int_{R^3} X^a_{\gamma}({\vec x}) 
A_a({\vec x}) d^3x, 
\label{eq:ha}\\
 X^a_{\gamma}({\vec x}) & := & 
\oint_{\gamma} ds \delta^3({\vec {\gamma}} (s), {\vec x}){\dot{\gamma}}^a,
\label{eq:x}                                
\end{eqnarray}
where $s$ is a parametrization of the loop $\gamma$, 
$s\in [0,2\pi]$.
$X^a_{\gamma}({\vec x})$ 
is called the form factor of $\gamma$. Its Fourier transform
is 
\begin{eqnarray}
X^a_{\gamma}({\vec k}) & : = & 
   {1\over 2\pi^{3\over 2}}\int_{R^3}d^3x X^a_{\gamma}({\vec x})
e^{-i{\vec k}\cdot {\vec x}}
\nonumber \\
& = & {1\over 2\pi^{3\over 2}}\oint_{\gamma} ds{\dot{\gamma}}^a(s) 
         e^{-i{\vec k}\cdot {\vec \gamma(s)}}.
\label{eq:xk}
\end{eqnarray}
The Gaussian smeared form factor \cite{lingrav} is defined as
\begin{equation}
 X^a_{\gamma_{(r)}}({\vec x}) := 
\int_{R^3} d^3y f_r({\vec y}-{\vec x})
            X^a_{\gamma}({\vec y}) 
  =  \oint_{\gamma} ds f_r({\vec \gamma (s)}- {\vec x}){\dot{\gamma}}^a(s)
\label{eq:xr}
\end{equation}
where
\begin{equation}
f_r({\vec x}) = {1\over 2\pi^{3\over 2}r^3}e^{-x^2\over 2r^2} 
\;\;\;x:=|{\vec x}| 
\end{equation}
approximates the Dirac delta function for small $r$.
The Fourier transform of the smeared form factor is
\begin{equation}
X^a_{\gamma_{(r)}}({\vec k}) = e^{-k^2r^2 \over 2}X^a_{\gamma}({\vec k})
\label{eq:xrk}
\end{equation}
and the smeared holonomy is defined as 
\begin{eqnarray}
H_{\gamma_{(r)}}(A)  &= &\exp i\int_{R^3} 
X^a_{\gamma_{(r)}}({\vec x}) A_a({\vec x}) d^3x 
\nonumber \\
 &= &\exp i\int_{R^3} X^a_{\gamma_{(r)}}(-{\vec k}) A_a({\vec k}) d^3k.
\label{eq:hr}
\end{eqnarray}
where $A_a({\vec k})$ is the Fourier transform of $A_a({\vec x})$.

We define
$\tilde{\alpha}_r$, $K_r$, ${\cal{HA}}_r$, $\overline{\cal{HA}}_r, \Delta_r$
as follows.

${\tilde \alpha}_r$ is the $r$-hoop class of $\alpha$ i.e.
 $\alpha,\beta$ define the same $r$-hoop iff 
$H_{\alpha_{(r)}} (A) =H_{\beta_{(r)}} (A)$
 for every $A_a(x)\in {\cal A} $. 
${\cal {HG}}_r$ is the group generated by all $r$-hoops ${\tilde \alpha}_r$ 
where
group multiplication is $r$-hoop composition i.e. 
\begin{equation}
{\tilde \alpha}_r\circ {\tilde \beta}_r := 
({\widetilde{\alpha\circ \beta}})_r.
\label{eq:rhgrp}
\end{equation}
Note that the above definition is consistent because, from
(\ref{eq:hr}) and the definition of $r$-hoop equivalence, it follows that
\begin{equation}
H_{\alpha_{(r)}} (A) H_{\beta_{(r)}} (A)
=H_{({\alpha\cdot\beta})_{(r)}}(A)
\end{equation}
Note that from (\ref{eq:rhgrp}), it follows that the identity element of 
${\cal {HG}}_r$ is ${\tilde o}_r$ and that 
$({\tilde \alpha}_r)^{-1}= {\widetilde {\alpha^{-1}}}_r$.

${\cal{HA}}_r$ 
is the abelian Poisson bracket algebra of the 
$r$-holonomies, $H_{\alpha_{(r)}} (A)$, $A_a \in {\cal A}$,
$\alpha \in {\cal L}_{x_0}$. 

Recall that
with the product law defined in section 2, all elements of 
${\cal FL}_{x_0}$ are expressible as complex linear combinations of 
elements of ${\cal L}_{x_0}$. We define  the  2 sided ideal of 
$K_r \in {\cal FL}_{x_0}$, through
\begin{equation}
\sum_{i=1}^N a_i \alpha_i \in K_r \;{\rm iff}\;\;
\sum_{i=1}^N a_i H_{{\alpha_i}_{(r)}}(A) = 0\;  {\rm for\; every} \;
A_a(x) \in {\cal A},
\end{equation}
where $a_i$ are complex numbers. 
The $K_r$ equivalence class of $\alpha$ is denoted by 
$[\alpha ]_r$.
It can be seen that, 
as abstract algebras, ${\cal{HA}}_r$  and ${\cal FL}_{x_0}/K_r $
are isomorphic. 

It can be checked that the relation $*_r$ defined on 
${\cal{HA}}_r$ by 
\begin{equation}
(\sum_{i=1}^N a_i [\alpha_i ]_r)^{*_r} := 
\sum_{i=1}^N a_i^*[ \alpha_i^{-1}]_r 
\label{eq:*r}
\end{equation}
is a $*$ relation. Note that from (\ref{eq:hr}), the complex 
conjugate of 
$H_{\alpha_{(r)}} (A)$ is $H_{{\alpha^{-1}}_{(r)}} (A)$ and hence the 
abstract $*_r$ relation just encodes the operation of complex conjugation on 
the algebra ${\cal{HA}}_r$.

Next we define the norm $|| \;\; ||_r$ as
\begin{equation}
||\sum_{i=1}^N a_i [\alpha_i ]_r||_r := 
\sup_{A\in {\cal A}} |\sum_{i=1}^N a_i H_{{\alpha_i}_{(r)}}(A)| .
\label{eq:rnorm}
\end{equation}
It is easily verified that $|| \;\; ||_r$ is indeed a norm on the 
$*$ algebra ${\cal{HA}}_r$ with $*$ relation defined by 
(\ref{eq:*r}).
 Completion of ${\cal{HA}}_r$ with respect to $|| \;\; ||_r$
gives the abelian $C^*$ algebra 
$\overline{\cal{HA}}_r$.

Next, we characterize the spectrum $\Delta_r$ of $\overline{\cal{HA}}_r$
as the space of all homomorphisms from 
${\cal {HG}}_r$ to $U(1)$.

Let $h \in \Delta_r$. Thus $h$ is a linear, multiplicative, continuous
 * homorphism from $\overline{\cal{HA}}_r$ to $\bf C$.
\begin{equation}
\Rightarrow h([\alpha ]_r) h([\alpha^{-1} ]_r) = h([o]_r).
\nonumber
\end{equation}
\begin{equation}
{\rm Choosing  \;\;}
\alpha=o, \;\; \Rightarrow h([o]_r)^2=h([o]_r) \;
\Rightarrow h([o]_r)=1.
\end{equation}
\begin{equation}
\Rightarrow h([\alpha^{-1} ]_r) = {1\over h([\alpha ]_r)}
          = h^*([\alpha ]_r) .
\label{eq:inverse}
\end{equation}
(\ref{eq:inverse})  implies that $|h([\alpha ]_r)|=1$
and this, coupled with the fact that ${\cal {HG}}_r$ is commutative, 
shows that every $h \in \Delta_r$ defines a homomorphism from 
${\cal {HG}}_r$ to $U(1)$.

Conversely, let $h$ be a homomorphism from 
${\cal {HG}}_r$ to $U(1)$. Its action can be extended by linearity to
elements of ${\cal{HA}}_r$ so that 
$h(\sum_{i=1}^N a_i [\alpha_i ]_r) := \sum_{i=1}^N a_i h([\alpha_i ]_r)$.
\footnote{$h$ can be defined on $[\alpha ]_r$ because $K_r$ equivalence 
subsumes $r$-hoop equivalence.} It is also easy to see that 
$h([\alpha^{-1} ]_r)= h^*([\alpha ]_r)$. These properties and the fact that
$h$ is a homomorphism from 
${\cal {HG}}_r$ to $U(1) \subset \bf C$, imply that $h$ is a linear,
multiplicative, * homomorphism from ${\cal{HA}}_r$ to $\bf C$.

Finally we show that $h$ extends to a continuous homomorphism
on $\overline{\cal{HA}}_r$. From \cite{ai} it follows that for 
$\alpha_i \in {\cal L}_{x_0}, \; i=1..n$, there exist strongly 
independent $\beta_j,\; j=1..m$ such that each $\alpha_i$ is the 
composition of some of the $\{\beta_j\}$. From this fact and Lemma 2 of 
Appendix A1, it can be shown that, for a given 
$\sum_{i=1}^N a_i [\alpha_i ]_r \in {\cal{HA}}_r$ and any $\delta >0$,
there exists $A^{(a_i,\delta,r)}_a \in {\cal A}$ such that 
\begin{equation}
|\sum_{i=1}^N a_i \big( h([\alpha_i ]_r)
-H_{\alpha_{i(r)}}(A^{(a_i,\delta,r)})\big)| < \delta .
\label{eq:aidelta}
\end{equation}
From (\ref{eq:aidelta}), it is straightforward to show that 
\begin{equation}
|\sum_{i=1}^N a_i \big( h([\alpha_i ]_r)| \leq 
\sup_{A\in {\cal A}} |\sum_{i=1}^N a_i H_{{\alpha_i}_{(r)}}(A)| .
=||\sum_{i=1}^N a_i [\alpha_i ]_r||_r .
\label{eq:hcont}
\end{equation}
Since ${\cal{HA}}_r$ is dense in $\overline{\cal{HA}}_r$, (\ref{eq:hcont})
implies that $h$ can be extended to a 
continuous (linear, multiplicative) homorphism from 
$\overline{\cal{HA}}_r$ to $\bf C$.

Thus $\Delta_r$ can be identified with the set of all homomorphisms from
${\cal {HG}}_r$ to $U(1)$.

\subsection*{3b. The isomorphism $I_r$}
We show that\\
\noindent (i) $K=K_r$ : Let 
\begin{equation}
\sum_{i=1}^N a_i H_{\alpha_i}(A) = 0\;  {\rm for\; every} \;
A_a(x) \in {\cal A}.
\label{eq:kdef}
\end{equation}
From Lemma 3 of Appendix A1, given
$A_a \in {\cal A}$, there exists $A_{a(r)} \in {\cal A}$ such that 
\begin{equation}
\sum_{i=1}^N a_i H_{\alpha_{i(r)}}(A) =
\sum_{i=1}^N a_i H_{\alpha_i}(A_{(r)}).
\label{eq:kkr}
\end{equation}
(\ref{eq:kdef}) and (\ref{eq:kkr}) imply that $K \subset K_r$.

Let 
\begin{equation}
\sum_{i=1}^N a_i H_{{\alpha_i}_{(r)}}(A) = 0\;  {\rm for\; every} \;
A_a(x) \in {\cal A}.
\end{equation}
$\Rightarrow$ Given $A_a, B_a \in {\cal A}$,
\begin{equation}
|\sum_{i=1}^N a_i H_{\alpha_i}(A)|= 
|\sum_{i=1}^N a_i H_{\alpha_i}(A) -\sum_{i=1}^N a_i H_{\alpha_{i(r)}}(B)|.
\end{equation}
Choose, $B_a = A^{\epsilon}_a$ where $A^{\epsilon}_a$ is defined          
in Lemma 1, A1. \\
\noindent
Then \\
\noindent
 $|\sum_{i=1}^N a_i H_{\alpha_i}(A)|\leq \sum_{i=1}^N |a_i|\epsilon$ for every 
$\epsilon > 0$. $\Rightarrow\sum_{i=1}^N a_i H_{\alpha_i}(A)=0$ and 
hence, $K_r\subset K$.

Thus $K=K_r$, $[\alpha ]= [\alpha ]_r$ and
${\tilde{\alpha}}={\tilde{\alpha}}_r$ 

\noindent (ii)$ ||\sum_{i=1}^N a_i [\alpha_i ]||
               = ||\sum_{i=1}^N a_i [\alpha_i ]_r||_r$: \\
\noindent
Let
$||\sum_{i=1}^N a_i [\alpha_i ]_r||_r =c_r$. 
Then 
$c_r \geq |\sum_{i=1}^N a_i H_{\alpha_{i(r)}}(A)|$ for every
$A_a \in {\cal A}$. Further,  for every $\tau >0$ there exists 
${}^{(\tau )}A_{a}\in {\cal A}$  such that 
$c_r -|\sum_{i=1}^N a_i H_{\alpha_{i(r)}}({}^{(\tau )}A)|\leq \tau$.Then, 
from Lemma 3, A1, there exists ${}^{(\tau )}A_{a(r)}\in {\cal A}$ such that 
\begin{equation}
0\leq c_r- \sum_{i=1}^N a_i H_{\alpha_i}({}^{(\tau )}A_{(r)}) \leq \tau.
\label{tau}
\end{equation}
\begin{equation}
\Rightarrow \sup_{A\in {\cal A}} |\sum_{i=1}^N a_i H_{\alpha_i}(A)|
\geq c_r \;\Rightarrow ||\sum_{i=1}^N a_i [\alpha_i ]||\geq
||\sum_{i=1}^N a_i [\alpha_i ]_r||_r
\end{equation}

Let
$||\sum_{i=1}^N a_i [\alpha_i ]|| =c$. Then 
for every $\tau >0$ there exists 
${}^{({\tau\over 2})}A_{a}\in {\cal A}$  such that 
\begin{equation}
c-|\sum_{i=1}^N a_i H_{\alpha_i}({}^{({\tau\over 2})}A)|\leq {\tau\over 2}.
\label{eq:tauhalf}
\end{equation}
From Lemma 1 A1, there exists ${}^{({\tau\over 2})}A_a^\epsilon \in {\cal A}$ 
such that 
\begin{eqnarray}
|H_{\alpha_{i(r)}}({}^{({\tau\over 2})}A^{\epsilon}) - 
H_{\alpha_{i}}({}^{({\tau\over 2})}A)| & \leq & \epsilon.\nonumber \\
\Rightarrow
|\sum_{i=1}^N a_i\big(H_{\alpha_{i(r)}}({}^{({\tau\over 2})}A^{\epsilon}) - 
H_{\alpha_{i}}({}^{({\tau\over 2})}A)\big)| & \leq & 
\sum_{i=1}^N |a_i|\epsilon.
\label{eq:epstau}
\end{eqnarray}
Choose $0<\epsilon < {\tau\over 2\sum_{i=1}^N |a_i|}$. From 
(\ref{eq:tauhalf}) and (\ref{eq:epstau})
it follows that, for every $\tau > 0$, 
\begin{equation}
c-|\sum_{i=1}^N a_i H_{\alpha_{i(r)}}({}^{({\tau\over 2})}
A^{\epsilon})|< \tau.
\end{equation}
\begin{equation}
\Rightarrow c \leq \sup_{A\in {\cal A}} |\sum_{i=1}^N a_i 
H_{\alpha_{i(r)}}(A)|
\; \Rightarrow 
||\sum_{i=1}^N a_i [\alpha_i ]||
               \leq ||\sum_{i=1}^N a_i [\alpha_i ]_r||_r.
\end{equation}

Thus, for any finite $N$, $||\sum_{i=1}^N a_i [\alpha_i ]||
               = ||\sum_{i=1}^N a_i [\alpha_i ]_r||_r$

From (i) and (ii) it follows that the structures 
$K_r$, $[\alpha_r]$, $\tilde{\alpha}_r$, ${\cal HG}_r$, $*_r$, 
${\cal{HA}}_r$, 
$\overline{\cal{HA}}_r, 
\Delta_r$ are isomorphic to 
$K$, $[\alpha]$, $\tilde{\alpha}$, ${\cal HG}$, *, ${\cal{HA}}$, 
$\overline{\cal{HA}}, 
\Delta$.

Thus a $C^*$ isomorphism 
$I_r:{\overline{\cal{HA}}} \rightarrow {\overline{\cal{HA}}}_r$  exists 
such that 
\begin{equation}
I_r(\sum_{i=1}^N a_i [\alpha_i ])= \sum_{i=1}^N a_i [\alpha_i ]_r .
\label{eq:irha}
\end{equation}
$I_r$ defines a natural 
1-1 map from $\Delta$ to $\Delta_r$ (which we shall also 
call $I_r$). Given the * isomorphism $h\in \Delta$, from 
${\overline{\cal{HA}}}$ to $\bf C$, its image is $h_r\in \Delta_r$
where 
\begin{equation}
h_r( \sum_{i=1}^N a_i [\alpha_i ]_r):= h (\sum_{i=1}^N a_i [\alpha_i ]_r).
\label{eq:irdelta}
\end{equation}

Note that if $h_r$ is
defined by some smooth, non-flat 
$A_a\in {\cal A}$ then {\em it is not true} 
that  $h$ is  
associated with (the gauge equivalence class of) the same connection.

\section*{4. Fock representation from  ${\overline{\cal{HA}}}_r$}

The standard Fock space vacuum expectation value 
 restricted to  ${\cal{HA}}_r$ defines the Fock PLF on ${\cal{HA}}_r$ as
\begin{equation}
\Gamma_F (\sum_{i=1}^N a_i [\alpha_i ]_r):= 
\sum_{i=1}^N a_i \exp -(\int {d^3k \over k} 
            | X^a_{\alpha_{i(r)}}({\vec k})|^2).
\label{eq:gammaf}
\end{equation}

Since ${\cal{HA}}_r$ is  a proper subalgebra of the standard 
Weyl algebra for 
$U(1)$ theory, it is not clear that its quantization 
(through the GNS construction based  on the Fock PLF) 
reproduces the full Fock space. We 
prove that the full Fock space is indeed obtained.

Let the GNS Hilbert space (based on $\Gamma_F$) be $\cal H$. Let 
$\cal D$ be the linear subspace of $\cal H$ spanned by elements of the 
form ${\hat H}_{\alpha_{(r)}} \Omega, \; \alpha\in {\cal L}_{x_0}$ 
where $\Omega$ is the GNS vacuum.
It can be seen that $\cal D$ is dense in $\cal H$. $\cal D$ is 
naturally embedded in the Fock space, $\cal F$, through the map 
$U:{\cal D}\rightarrow {\cal F}$ defined by 
\begin{equation}
 U (\Omega ) = |0> \;\;{\rm and}\;\;
 U(\sum_{i=1}^N a_i {\hat H}_{\alpha_{i(r)}} \Omega )
  =\sum_{i=1}^N a_i\exp i\int_{R^3}
X^a_{\alpha_{i(r)}}({\vec x}) {\hat A}_a({\vec x}) d^3x |0> .
\label{eq:udef}
\end{equation}
Here ${\hat A}_a$ is the standard Fock space operator valued distribution
at $t=0$
\begin{equation}
{\hat A}_a({\vec x})= {1\over 2\pi^{3\over 2}}
\int {d^3k\over {\sqrt{k}}} (e^{i{\vec k}\cdot {\vec x}}{\hat a}_a({\vec k})
         +e^{-i{\vec k}\cdot {\vec x}}{\hat a}^{\dagger}_a({\vec k})
\label{eq:defa}
\end{equation}
where 
\begin{equation}
{\hat a}_a({\vec k}) k^a= 0,  \;\;\; 
[{\hat a}_a({\vec k}),{\hat a}^{\dagger}_b({\vec l})] 
= \delta_{ab} \delta ({\vec k},{\vec l}).
\label{eq:comma}
\end{equation}

By construction, $U$ is a unitary map and can be uniquely extended to
$\cal H$ so that it embeds ${\cal H}$ in $\cal F$.
We show that Cauchy limits of states in $U({\cal D})$ span a dense set
in $\cal F$ - this suffices to show that the entire Fock space is indeed
obtained, i.e. that $U({\cal H}) ={\cal F}$.

Define the `occupation number' states 
\begin{equation}
|\phi, p> := \int d^3k_1..d^3k_p \phi^{a_1..a_p}({\vec k}_1..{\vec k}_p)
{\hat a}^{\dagger}_{a_1}({\vec k}_1)..{\hat a}^{\dagger}_{a_p}({\vec k}_p)
|0>.
\label{eq:phip}
\end{equation}
$\phi^{a_1..a_p}({\vec k}_1..{\vec k}_p)$ (with $p$ a positive integer)
 is such that \\
\noindent
(a) $\int d^3k_i|\phi^{a_1..a_p}({\vec k}_1,..,{\vec k}_i,..,{\vec k}_p)|^2
< \infty$ and $\phi^{a_1..a_p}({\vec k}_1..{\vec k}_p)$ falls of 
faster than any inverse power of $k_i$ as 
$k_i\rightarrow \infty,\; {\vec k}_{j\neq i}$ fixed.\\
\noindent
(b) $\phi^{a_1..a_i..a_p}({\vec k}_1,..,{\vec k}_i,..,{\vec k}_p)
(k_i)_{a_i}=0$ i.e. it is transverse.\\
\noindent
(c) it is symmetric under interchange of $(a_i,{\vec k}_i)$ with 
$(a_j,{\vec k}_j)$ for all $i,j=1..p$. 

$|\phi, p>$ for all $p$ together with $|0>$, 
span a dense set, ${\cal D}_0 \in {\cal F}$.

Given 2 vectors ${\vec x}, {\vec v}$, define the operator
\begin{equation}
{\hat O}_{{\vec x}, {\vec v}}:=
{i \over 2\pi^{3\over 2}}\int {d^3k \over {\sqrt{2}} k} 
e^{i{\vec k}\cdot {\vec x}}e^{-k^2r^2\over 2}({\vec v} \times {\vec k})^a 
({\hat a}_a({\vec k})
         +{\hat a}^{\dagger}_a(-{\vec k})).
\label{eq:defo}
\end{equation}
As argued in Appendix A2, states of the form 
$|\psi_{\{{\vec x}_i, {\vec v}_i\}}> :=
\prod_{i=1}^p{\hat O}_{({\vec x}_i, {\vec v}_i)}|0>,\; p=1,2..$ together with 
$|0>$ span ${\cal D}_0$.

Our proof  that $\psi_{\{{\vec x}_i, {\vec v}_i\}}\in U({\cal H})$
is as follows.\\
\noindent (i) Note
that 
$\psi_{\{{\vec x}_i, {\vec v}_i\}}\in {\cal D}_0$. \\
\noindent (ii) Let $\gamma^{\{m, {\vec x}, {\vec n}\}}$ be a circular loop 
of radius $\epsilon_m :={1\over 2^m}$ ($m$ is a positive integer),
 centred at ${\vec x}$ 
and let its plane have unit normal
$\vec n$ \footnote{Although $\gamma^{\{m, {\vec x}, {\vec n}\}}$
 is not in ${\cal L}_{x_0}$, the loop formed 
by joining  
$\gamma^{\{m, {\vec x}, {\vec n}\}}$ to the base point $x_0$ and
retracing, is. We shall continue to denote this loop, 
which represents the same hoop,  
by 
$\gamma^{\{m, {\vec x}, {\vec n}\}}$.}.
The image of ${\hat H}_{\gamma_{(r)}^{\{m, {\vec x}, {\vec n}\}}}$ on 
$U({\cal D})$ is 
$\exp(i\int 
X^a_{\gamma_{(r)}^{\{m, {\vec x}, {\vec n}\}}}({\vec y}) {\hat A}_a({\vec y}) 
d^3y)$. Define 
\begin{equation}
{\hat O}_{{\vec x},{\vec n}, m}:= 
{e^{i\int 
X^a_{\gamma_{(r)}^{\{m, {\vec x}, {\vec n}\}}}({\vec y}) {\hat A}_a({\vec y}) 
d^3y}- 1\over i\pi \epsilon_m^2}.
\label{eq:defom}
\end{equation}
The formal limit of ${\hat O}_{{\vec x},{\vec n}, m}$ as 
$m\rightarrow \infty$
is ${\hat O}_{{\vec x}, {\vec n}}$. We show below that 
${\hat O}_{{\vec x},{\vec n}, m}
|\psi >$, $|\psi> \in {\cal D}_0$,
\footnote{Note that since ${\hat O}_{{\vec x},{\vec n}, m}$ 
are bounded operators defined on the entire 
Fock space, ${\hat O}_{{\vec x},{\vec n}, m}$ are  well defined on 
${\cal D}_0$.}
  form a Cauchy sequence
with limit ${\hat O}_{{\vec x},{\vec n}}
|\psi >$. Then, choosing  $|\psi >= |0>$, we see that  
${\hat O}_{{\vec x},{\vec n}}|0>$ is in the completion of $U({\cal D})$.\\

\noindent (iii) 
From (i) above, ${\hat O}_{{\vec x},{\vec n}}|0> \in {\cal D}_0$. 
We can repeat the argument in (ii) above to conclude that  
  ${\hat O}_{{\vec x}_2,{\vec n}_2}
{\hat O}_{{\vec x}_1,{\vec n}_1}|0>$
 is obtained as the Cauchy limit of the states
${\hat O}_{{\vec x}_2,{\vec n}_2, m}{\hat O}_{{\vec x}_1,{\vec n}_1}|0>$. 
Iterating this argument we see that 
$|\psi_{\{{\vec x}_i, {\vec n}_i\}}>, \; |{\vec n}_i|=1 $
is in the 
completion of $U({\cal D})$. Finally, set ${\vec v}_i:= v_i{\vec n}_i$,
where $v_i$ are real numbers. Then it follows that 
$|\psi_{\{{\vec x}_i, {\vec v}_i\}}>=
(\prod_{i=1}^p v_i)|\psi_{\{{\vec x}_i, {\vec n}_i\}}>$ and hence
that $|\psi_{\{{\vec x}_i, {\vec v}_i\}}>\in U({\cal H})$.

Thus, it remains to show (see (ii) above) that:

\noindent  Given $\psi \in {\cal D}_0$, 
${\hat O}_{{\vec x}, {\vec n}}$ as defined in (\ref{eq:defo}) and
${\hat O}_{{\vec x},{\vec n}, m}$ as defined in (\ref{eq:defom}), 
\begin{equation}
\lim_{m\rightarrow \infty} 
||{\hat O}_{{\vec x}, {\vec n}}-{\hat O}_{{\vec x},{\vec n}, m}|\psi> ||
=0
\label{eq:cauchy}
\end{equation}

\noindent {\em Proof}:
Let 
\begin{equation}
{\hat D}:= \int 
X^a_{\gamma_{(r)}^{\{m, {\vec x}, {\vec n}\}}}({\vec y}) {\hat A}_a({\vec y}) 
d^3y - \pi \epsilon_m^2{\hat O}_{({\vec x}, {\vec n})}. 
\label{eq:defD}
\end{equation}
Thus 
\begin{equation}
{\hat O}_{{\vec x}, {\vec n}, m} = 
{e^{i\pi\epsilon_m^2 {\hat O}_{{\vec x}, {\vec n}}+ i{\hat D}} -1
\over i \pi \epsilon_m^2}.
\label{eq:omd}
\end{equation}
Since
 both $e^{i\pi \epsilon_m^2 {\hat O}_{{\vec x}, {\vec n}}}$
and $e^{i{\hat D}}$ are commuting elements of the standard Weyl algebra,
\begin{equation}
{\hat O}_{{\vec x}, {\vec n}, m}= 
{e^{i\pi \epsilon_m^2 {\hat O}_{{\vec x}, {\vec n}}}
e^{i{\hat D}}- 1\over i\pi\epsilon_m^2}
\end{equation}
\begin{eqnarray}
\Rightarrow 
||{\hat O}_{{\vec x}, {\vec n}}-{\hat O}_{{\vec x},{\vec n}, m}|\psi> ||
\;\;\;\;\;\;\;\;\;\;\;\;\;\;\;\;\;\;\;\;\;\;\;\;\;\;\;\;\;\;\;
\nonumber \\
=   
||{e^{i\pi \epsilon_m^2 {\hat O}_{{\vec x}, {\vec n}}}
(e^{i{\hat D}} -1)\over i\pi\epsilon_m^2} |\psi >
+
\big({e^{i\pi \epsilon_m^2 {\hat O}_{{\vec x}, {\vec n}}}- 1
  \over i\pi\epsilon_m^2} - {\hat O}_{{\vec x}, {\vec n}}\big)|\psi >|| 
\nonumber \\
  \leq  
||\big({e^{i\pi \epsilon_m^2 {\hat O}_{{\vec x}, {\vec n}}}- 1
  \over i\pi\epsilon_m^2} - {\hat O}_{{\vec x}, {\vec n}}\big)|\psi >||
+  
||{e^{i\pi \epsilon_m^2 {\hat O}_{{\vec x}, {\vec n}}}
(e^{i{\hat D}} -1)\over i\pi\epsilon_m^2} |\psi >||
\label{eq:ineqdo}
\end{eqnarray}
From Lemma 2 of Appendix A2, ${\hat O}_{{\vec x}, {\vec n}}$ is a densely 
defined symmetric operator on ${\cal D}_0$ and admits self adjoint extensions.
Hence, from \cite{rs}, the first term in (\ref{eq:ineqdo}) 
vanishes in the $\epsilon_m \rightarrow 0$ limit. Further, since 
$e^{i\pi \epsilon_m^2 {\hat O}_{{\vec x}, {\vec n}}}$ is a unitary operator,
we have
\begin{equation}
||{e^{i\pi \epsilon_m^2 {\hat O}_{{\vec x}, {\vec n}}}
(e^{i{\hat D}} -1)\over i\pi\epsilon_m^2} |\psi >||
= ||{(e^{i{\hat D}} -1)\over i\pi\epsilon_m^2} |\psi >||.
\end{equation}
But 
\begin{equation}
||{(e^{i{\hat D}} -1)\over i\pi\epsilon_m^2} |\psi >||^2
= -\left(<\psi |{(e^{i{\hat D}} -1)\over \pi^2\epsilon_m^4} |\psi > 
+<\psi |{(e^{-i{\hat D}} -1)\over \pi^2\epsilon_m^4} |\psi >\right) .
\label{eq:eid0}
\end{equation}
From Lemma 3, A2 and (\ref{eq:eid0}), 
$||{(e^{i{\hat D}} -1)\over i\pi\epsilon_m^2} |\psi >||\rightarrow 0$ 
as $\epsilon_m\rightarrow 0$ and then (\ref{eq:ineqdo}) implies 
(\ref{eq:cauchy}).

Thus we have 
shown above that the GNS representation of ${\cal{HA}}_r$ on
the GNS Hilbert space $\cal{H}$, 
is unitarily equivalent to the standard
Fock representation on ${\cal F}= L^2 ({\cal S}^{\prime}, d\mu_G )$ 
(${\cal S}^{\prime}$ denotes the appropriate space of tempered 
distributions and $\mu_G$ is the standard Gaussian measure with 
covariance ${1\over 2} (-\nabla^2)^{-{1\over 2}}$ \cite{rsqft})
 via the unitary map $U$. 

The action of the smeared electric field operator,
${\hat E}({\vec f}):= \int d^3xf_a({\vec x}){\hat E}^a({\vec x})$,
 on $\psi \in {\cal C}_F\subset L^2 ({\cal S}^{\prime}, d\mu_G )$
is written in the standard way \cite{rsqft} as
\begin{equation}
{\hat E}({\vec f})\psi = e\int d^3x\big(
     -i f_a({\vec x}){\delta\over \delta A_a({\vec x})}
       + i( (-\nabla^2)^{{1\over 2}}f^a({\vec x}))A_a({\vec x})  
                                    \big) \psi.
\label{eq:defef}
\end{equation}
Here $f_a({\vec x})$ is real, divergence free, smooth and of rapid decrease,
and ${\cal C}_F\subset L^2 ({\cal S}^{\prime}, d\mu_G )$ is the standard 
dense domain of cylindrical functions appropriate to Fock space.
The smeared electric field operator on $\cal{H}$
is defined as 
the unitary image of ${\hat E}({\vec f})$ by $U^{-1}$ i.e. for 
$\psi \in U^{-1}({\cal C}_F)\subset \cal{H}$
\begin{equation}
{\hat E}({\vec f})\psi = U^{-1}e\int d^3x\big(
     -i f_a({\vec x}){\delta\over \delta A_a({\vec x})}
       + i( (-\nabla^2)^{{1\over 2}}f^a)({\vec x})A_a({\vec x})  
                                    \big) U\psi.
\label{eq:defefu}
\end{equation}
With this action, ${\hat E}({\vec f})$ is densely defined on the dense domain
$U^{-1}({\cal C}_F)\subset\cal{H} $, and just like its unitary 
image on ${\cal C}_F$, admits a unique self adjoint extension.

\section*{5. Induced representations through $I_r$.}

It can be verified that $I_r$ is a topological homorphism from 
$\Delta$ to $\Delta_r$ (where $\Delta$ and $\Delta_r$ are equipped with 
their Gel'fand topologies). Hence, $I_r$ defines a measurable 
isomorphism $I_r:{\cal B}\rightarrow {\cal B}_r$ where 
${\cal B}$ and ${\cal B}_r$ are the Borel sigma algebras associated with 
$\Delta$ and $\Delta_r$ respectively.
Any regular Borel measure $\mu$ on $\Delta$ induces a regular Borel 
measure $\mu_r$ on $\Delta_r$, with $\mu_r := \mu I_r^{-1}$. It follows
that $I_r$ defines a unitary map $U_r$ from $L^2 (\Delta, d\mu )$
to $L^2 (\Delta_r, d\mu_r )$. 

$U_r$ can be explicitly defined through its action on the dense set 
${\cal C}\in L^2 (\Delta, d\mu )$, 
of cylindrical functions (cylindrical functions in the 
context of ${\overline{\cal{HA}}}$ have been defined in section 2).
Denote the dense set of cylindrical functions in 
$L^2 (\Delta_r, d\mu_r )$ by ${\cal C}_r$ 
\footnote{Cylindrical functions are of the form
$\psi_{\{[\alpha_i ]_r\}}(h):= \psi (h([\alpha_1 ]_r)..h([\alpha_n ]_r))$,
 for $\alpha_i \in {\cal L}_{x_0}, i=1..n$, $h\in \Delta_r$, 
and  they span ${\cal C}_r$.}.
$U_r$ maps $\cal C$ to ${\cal C}_r$ through
\begin{equation}
U_r( \psi_{\{[\alpha_i ]\}} ) =\psi_{\{[\alpha_i ]_r\}}.
\label{eq:ur1}
\end{equation}
It also follows that 
\begin{equation}
U_r {\hat H}_{\alpha} U_r^{-1} = {\hat H}_{\alpha_{(r)}}.
\label{eq:ur2}
\end{equation}
Thus $I_r$ induces a representation of $\overline{\cal{HA}}_r$
from a representation of $\overline{\cal{HA}}$. 
In section 5a, we induce a Haar like representation of $\overline{\cal{HA}}_r$
from the Haar representation of $\overline{\cal{HA}}$. 

Since the image of $I_r$ restricted to $\cal{HA}$ is ${\cal{HA}}_r$,
$I_r$ (or $I_r^{-1}$) can also be used to induce representations of 
$\cal{HA}$ from those of ${\cal{HA}}_r$ and vice versa. 
In 
section 5b, we induce a Fock like representation of 
$\cal{HA}$ from the Fock representation 
of ${\cal{HA}}_r$. The elements of 
${\cal{HA}}_r$ define a dense subspace of $\cal{H}$ through the GNS 
construction and a map $U_r$ is defined through (\ref{eq:ur1}) and 
(\ref{eq:ur2}). $U_r^{-1}$ induces a Fock like representation of $\cal{HA}$.

\subsection*{5a. Haar  representation of ${\overline{\cal{HA}}}_r$} 
We denote both the Haar measure on $\Delta$ as well as its image on 
$\Delta_r$ by $d\mu_H$.
The induced PLF (corresponding to $d\mu_H$) 
on ${\overline{\cal{HA}}}_r$ is defined by 
\begin{eqnarray}
\Gamma(\alpha )& =& 1\; {\rm if}\; \tilde{\alpha_r} =\tilde{o_r}\nonumber \\
               & = & 0\;\;\;{\rm otherwise}.
\label{eq:rhaarplf}
\end{eqnarray}
From (\ref{eq:ur1}) and (\ref{eq:ur2}) it follows that 
${\hat H}_{\alpha_{(r)}}$ are represented by unitary operators on 
$\psi \in L^2 (\Delta_r, d\mu_H )$ by 
\begin{equation}
({\hat H}_{\alpha_{(r)}}\psi) (h) = h([\alpha ]_r)\psi (h), \;\; h\in\Delta_r.
\label{eq:ur3}
\end{equation}

We construct electric field 
operators on $L^2 (\Delta_r, d\mu_H )$
as unitary images of appropriate electric field operators on 
$L^2 (\Delta_, d\mu_H )$ as follows.
Define the classical Gaussian smeared electric field  as 
\begin{equation}
E^a_r({\vec x}) := 
\int d^3y f_r({\vec y}-{\vec x})
            E^a({\vec y}) 
\label{eq:defer}
\end{equation}
where $f_r$ has been defined in section 3.
Given $\psi_{\{[\alpha_i ]\}}  \in  {\cal C} \subset L^2(\Delta, d\mu_H)$
it can be checked that 
\begin{equation}
({\hat E}^a_r({\vec x})\psi_{\{[\alpha_i ]\}})(h) 
=  e \sum_{i=1}^nX^a_{\alpha_{i(r)}}({\vec x}) h([\alpha_i ]) 
{\partial \psi_{\{[\alpha_i ]\}}\over \partial h([\alpha_i ])}.
\label{eq:defhater}
\end{equation}
The methods of \cite{alproj} can be used
to show that ${\hat E}^a_r({\vec x})$ 
is  essentially self adjoint on $\cal C$. Note that
(\ref{eq:defhater}) implies that 
\begin{equation}
[{\hat E}^a_r({\vec x}), {\hat H}_{\alpha}] =
         e X^a_{\alpha_{(r)}}({\vec x}){\hat H}_{\alpha}.
\label{eq:comerh}
\end{equation}
The unitary image of (\ref{eq:comerh}) is 
\begin{equation}
[U_r{\hat E}^a_r({\vec x})U_r^{-1}, {\hat H}_{\alpha_{(r)}}] =
        e X^a_{\alpha_{(r)}}({\vec x}){\hat H}_{\alpha_{(r)}}.
\label{eq:comuerh}
\end{equation}
Denote the classical counterpart of $U_r{\hat E}^a_r({\vec x})U_r^{-1}$
by $F^a({\vec E})$. Then (\ref{eq:comuerh}) provides a quantum 
representation of the classical Poisson bracket,
\begin{equation}
\{F^a({\vec E})
,  H_{\alpha_{(r)}}(A)\} =
        -ie X^a_{\alpha_{(r)}}({\vec x}) H_{\alpha_{(r)}}(A).
\label{eq:Fpb}
\end{equation}
Note that $\{E^a({\vec x})
,  H_{\alpha_{(r)}}(A)\} =
        -ie X^a_{\alpha_{(r)}}({\vec x}) H_{\alpha_{(r)}}(A)$. Hence, we
can consistently identify $F^a({\vec E})$ with $E^a({\vec x})$. 
Thus, 
$U_r{\hat E}^a_r({\vec x})U_r^{-1}= {\hat E}^a({\vec x})$
and from (\ref{eq:defhater}),
\begin{equation}
({\hat E}^a({\vec x})\psi_{\{[\alpha_i ]_r\}})(h) 
=  e \sum_{i=1}^nX^a_{\alpha_{i(r)}}({\vec x}) h([\alpha_i ]_r)
{\partial \psi_{\{[\alpha_i ]_r\}}(h)\over \partial h([\alpha_i ]_r)}.
\label{eq:defhate}
\end{equation}
Since $U_r$ is unitary, ${\hat E}^a({\vec x})$ is essentially self adjoint
on ${\cal C}_r\subset L^2 (\Delta_r, d\mu_H )$. 

To summarize: The induced Haar representation of 
${\overline{\cal{HA}}}_r$ provides a quantum representation of the classical 
Poisson bracket algebra of smeared holonomies $H_{\alpha_{(r)}}(A)$ and
(divergence free) electric field $E^a({\vec x})$.
${\hat H}_{\alpha_r}$ are represented by unitary operators through 
(\ref{eq:ur3}) and the {\em unsmeared} electric field operator, 
${\hat E}^a({\vec x})$,  is represented through  (\ref{eq:defhate})
as an essentially self adjoint operator
on the dense domain of cylindrical functions,
${\cal C}_r\subset L^2 (\Delta_r, d\mu_H )$. 
Note that ${\hat E}^a({\vec x})$ is a genuine operator as 
opposed to an operator valued distribution!

\subsection*{5b. Fock representation of $\cal{HA}$}
We denote  the `Fock' PLF on ${\cal{HA}}_r$ as well as its image on $\cal{HA}$
 by $\Gamma_F$.
Note that the  induced PLF 
on $\cal{HA}$ is defined by 
\begin{equation}
\Gamma_F (\sum_{i=1}^N a_i [\alpha_i ]):= 
\sum_{i=1}^N a_i \exp -(\int {d^3k \over k} 
            | X^a_{\alpha_{i(r)}}({\vec k})|^2).
\label{eq:gammafha}
\end{equation}

Since ${\hat H}_{\alpha_{(r)}}$ are represented as unitary operators on 
$\cal{H}$, it follows that 
\begin{equation}
{\hat H}_{\alpha}:= U_r^{-1} {\hat H}_{\alpha_{(r)}} U_r 
\label{eq:defhaf}
\end{equation}
are represented as unitary operators on $U_r^{-1}(\cal{H})$.

It remains to construct, following the strategy of section 5a, electric field
operators on $U_r^{-1}(\cal{H})$ as unitary images of appropriate 
electric field operators on $\cal{H}$.
On $U^{-1}({\cal C}_F) \subset \cal{H}$,
\begin{equation}
[{\hat E}({\vec f}), {\hat H}_{\alpha_{(r)}}] =
        e \int d^3x f_a({\vec x})
X^a_{\alpha_{(r)}}({\vec x}){\hat H}_{\alpha_{(r)}}.
\label{eq:comehr}
\end{equation}
\begin{equation}
\Rightarrow
[U_r^{-1}{\hat E}({\vec f})U_r, {\hat H}_{\alpha}] =
         e \int d^3x f_a({\vec x})
X^a_{\alpha_{(r)}}({\vec x}){\hat H}_{\alpha}.
\label{eq:comuehr}
\end{equation}
Define the classical function
\begin{equation}
E_r({\vec f}) := \int d^3x  f_a({\vec x})E^a_r({\vec x}),
\label{eq:erf}
\end{equation}
where $E^a_r({\vec x})$ is defined by (\ref{eq:defer}).
Since 
\begin{equation}
\{E_r({\vec f})
,  H_{\alpha}(A)\} =
        -ie \int d^3x f_a({\vec x})
X^a_{\alpha_{(r)}}({\vec x}) H_{\alpha}(A),
\label{eq:erhapb}
\end{equation}
we identify 
\begin{equation}
{\hat E}^a_r({\vec f}):= U_r^{-1}{\hat E}({\vec f})U_r .
\label{eq:5blast}
\end{equation}
To summarize: The induced Fock  representation of 
${\cal{HA}}$ provides a quantum representation of the classical 
Poisson bracket algebra of holonomies $H_{\alpha}(A)$ and
``Gaussian-smeared, smeared''
 electric fields $E_r({\vec f})$.
The `{\em unsmeared}' holonomy operators
${\hat H}_{\alpha}$ are represented by unitary operators through 
(\ref{eq:defhaf}) and 
${\hat E}({\vec f})$  is represented as a self adjoint operator
through  (\ref{eq:5blast}).
Note that the ``Gaussian smeared object''
 ${\hat E}_r({\vec x})$ is represented as an 
{\em operator valued distribution} (as opposed to a genuine operator)
on $U_r^{-1}(\cal{H})$.

\section*{6. Discussion}
\noindent {\em Preliminary remarks}: In this paper, representations of the 
Poisson algebra of $U(1)$  theory were constructed in 2 steps.
First, Hilbert space representations of the abelian Poisson algebra of 
configuration functions (i.e. functions of $A_a$) were constructed by 
specifying a PLF. Second, real functions of the conjugate electric field were 
represented by self adjoint operators on this Hilbert space. The Haar 
representation of ${\overline{\cal HA}}$ and its image on 
${\overline{\cal HA}}_r $ 
support a representation of the electric field wherein, formally,
\begin{equation}
{\hat E}^a({\vec x}) =-i {\delta \over \delta A_a({\vec x})}.
\label{eq:6.1}
\end{equation}
This action is not connected with Poincare invariance and it is not 
surprising that the resulting  representations of section 2 and 5a, are non
Fock representations.
On 
the Fock representation of ${\cal HA}_r $
\footnote{We remind the reader that we displayed a fairly rigorous
argument that the entire Fock space is obtained in such a representation
through the constructions of section 4 and Appendix A2. We reiterate
our belief that the formal equation (\ref{eq:formal}) can be rendered 
mathematically well defined in a more careful treatement.}
 and its image 
on ${\cal HA}$, equation (\ref{eq:6.1}) is incompatible with the 
requirement of self adjointness of the electric field operators. Their 
action necessarily contains a term dependent on the Gaussian measure 
(see (\ref{eq:defefu})) to ensure self adjointness. The 
choice of Gaussian 
measure is intimately associated with the properties of the 
de Alembertian, ${\partial^2 \over \partial t^2} - \nabla^2$, 
and hence with Poincare invariance.

A rephrasing of the above remarks which 
 brings them closer to the strategy of \cite{lingrav, extloop} is as follows. 
Given a representation in which (smeared or unsmeared) holonomies are 
represented by multiplication by unitary operators and the electric 
field acts, 
as in (\ref{eq:6.1}), purely by functional differentiation, 
the requirement of self adjointness of the electric field
operator determines the Hilbert space measure to be the Haar measure.
The self adjointness of electric field operators results in the Gaussian 
measure only if their action has a contribution dependent on the 
Gaussian measure.  Thus, to obtain the standard Fock representation or the
induced one of section 5b, the Gaussian measure and hence, Poincare 
invariance,
plays an essential and explicit role.

Note that this work concerns the `connection' representation of a theory
of a real $U(1)$ connection. In contrast \cite{lingrav} constructs the 
{\em loop} representation of a description of linearised gravity based on
a {\em self dual} connection \footnote{Note, however that the descripton 
reduces to one in terms of a triplet of abelian connections.}.
Despite these differences, there is also a certain amount of shared 
mathematical structure in our work and \cite{lingrav}. Therefore, the  
delineation of the structures involved in the construction of $U(1)$ theory
as spelt out in this paper, allows us to identify the role of the key 
structures in \cite{lingrav}. 

\noindent{\em Discussion of (I) and (II)}: We use the notation of 
\cite{lingrav} and \cite{extloop} when discussing those papers.
We first discuss (I). In \cite{lingrav} the action of the linearised metric 
variable in the loop representation is deduced, ultimately, from its 
action of the form `$-i{\delta\over \delta A_a^i}$' in the connection 
representation. The loop representation then becomes an `electric field'
type representation in which the magnetic field operator acts 
purely by functional 
differentiation with respect to the loop form factor.
Yet a Fock representation (of the positive and negative 
helicity gravitons)  results in apparent contradiction to our claims that 
such a representation cannot result without using Poincare invariance
explicitly.

The 
resolution of this apparent contradiction for the positive helicity graviton
sector seems to 
lie, in what appears at first sight, to be a mere mathematical nicety. 
In \cite{lingrav} the 
Gaussian measure contribution to the ${\hat B}^+$ operator is absorbed
(and hidden) in the rescaling of the wave function. Such a rescaling 
is permissible for finite dimensional systems but  results in a 
mathematically ill defined `measure' for the field theory in question.
In spite of the fact that for most applications this formal treatment 
suffices, it is crucial to realise, in the context of (I), that it hides the
role of Poincare invariance in constructing the Fock representation.
 To obtain a well defined (Gaussian) measure, the wave functions ((101) of
\cite{lingrav}) need to be rescaled  and a Gaussian measure term needs to be 
added  to the action of the ${\hat B}^+$ - this, of course, feeds 
explicit Poincare invariance back into the construction.

Note that this argument does not apply to the negative helicity sector.
There, 
the choice of self dual connection  results in the negative
helicity magnetic field operator, ${\hat B}^-$ being the same as 
the negative helicity annihilation operator. ${\hat B}^-$ is naturally 
represented as a functional derivative ((70) of \cite{lingrav}) and,
in this aspect, matches the standard Fock representation of the 
annihilation operator as a pure functional derivative term. The resulting 
representation is the Fock representation for negative helicity gravitons and
indeed,  for this sector,
 it seems that explicit Poincare invariance is not invoked.

Thus the Fock representation of linearised gravity 
seems to result partly due to explicit Poincare invariance (which is 
suppressed in \cite{lingrav} by a mathematically ill defined operation) and 
partly due to the use of self dual connections.

Considerations similar to those for the positive helicity gravitons 
also apply 
to the treatment of free Maxwell theory in \cite{extloop}. There, it is 
shown that the extended loop representation coincides, formally, with the 
electric field representation. Again, an ill defined measure is used and a
proper mathematical treatment restores the explicit role of Poincare 
invariance.

We turn now,  to a discussion of (II).
In the loop representation of \cite{lingrav}
the two sets of important operators are the magnetic field, ${\hat B}^{\pm}$,
and the linearised metric, ${\hat h}^{\pm}$. They are 
represented by functional differentiation and multiplication, on the 
representation space of functionals of loop form factors. This representation 
space supports the holonomies as operators. Indeed, the action of the 
${\hat B}^{\pm}$ operators is deduced from the fact that the classical 
magnetic flux is the lowest non trivial term in the expansion of the 
holonomy of a small loop ((57) of \cite{lingrav}). 
However, all these constructions are rendered formal because of the 
distributional nature of the loop form factor and the resulting 
divergence of the ground state functional. 
Therefore a regularization procedure is adopted wherein the loop form factors
are replaced by their Gaussian smeared, $r$- versions (see (\ref{eq:xr}))
and 
${\hat B}^{\pm}$, ${\hat h}^{\pm}$ are represented as functional 
differentiation and multiplication operators on the space of functionals 
of $r$-loop form factors.
An important question is: Are the holonomy operators or some 
regularized version thereof, represented on this space?

One may choose to ignore this question and simply postulate the action 
of ${\hat B}^{\pm}$, ${\hat h}^{\pm}$ in terms of $r$-loop form factors. Then
the primary configuration variables of the theory are $B^{\pm}$ and 
the construction does not seem to have much to do with loops and holonomies.
Since 
holonomies and loops are the primary objects in the loop approach
to  full-blown quantum gravity, an interpretation of the regularization 
which allows for the representation of holonomy operators is of 
interest. It seems to us that such an interpretation must regard
the $r$- form factor representation as an {\em approximation}
to the standard Fock representation, which becomes better as 
$r\rightarrow 0$. 

A precise formulation of such an interpretation in the context of $U(1)$ 
theory is provided by the induced Fock representation of section 5b. There,
the holonomy operators are represented on the Hilbert space and the 
magnetic field operators can be constructed by a ``shrinking of loop'' limit,
as the image of $U^{-1}{\hat O}_{{\vec x},{\vec n}}U$ via $U_r^{-1}$. 
That representation, although {\em not} the standard Fock representation,
is a good approximation to it for small $r$. The nature of the 
approximation is as follows. For sufficiently small $r$, the holonomies
$H_{\gamma}(A)$, the electric field $E^a({\vec x})$ 
and their Gaussian smeared counterparts,
$H_{\gamma_{(r)}}(A)$, $E^a_r({\vec x})$ approximate each other well.
An approximate Fock representation can be constructed in which the operators 
corresponding to 
$H_{\gamma}(A)$, $E^a_r({\vec x})$ act in the same way as the operators
corresponding to 
$H_{\gamma_{(r)}}(A)$, $E^a({\vec x})$
in the standard Fock representation.
This approximate Fock representation is the induced Fock representation 
of section 5b.

To summarize: The standard Fock representation 
for $U(1)$ theory is obtained only when 
the algebra of smeared holonomies is used  {\em and} explict Poincare
invariance is invoked. However, the role of Poincare invariance 
(or equivalently, the choice of PLF) seems to be more important than that of 
smeared loops. If the requirement of smeared loops is dropped, it is still
possible to construct an approximate Fock representation; but dropping 
Poincare invariance results in the non Fock representations of 
section 2 and 5a. 

\noindent {\em Comments}: \\
\noindent (i) The `area derivative' plays an important 
role in some approaches to loop quantum gravity \cite{carlolee,extloop}.
 Our construction of ${\hat O}_{{\vec x},{\vec n}}|\psi >$ 
(or its image in the induced Fock representation of 5b)
as a Cauchy limit 
is a rigorous realization of the area derivative in the context of
 Fock like representations. Note that the required limits do not 
exist in the Haar representation and hence the area derivative is ill defined
there.\\
\noindent (ii) As noted above, 
self duality of the connection plays a 
key role
in obtaining the 
(negative helicity)
graviton Fock representation without explicit recourse to 
Poincare invariance (see \cite{hel} for a detailed examination of the 
relation between self duality and helicity). 
However, recent efforts in loop quantum gravity 
use real (as opposed to self dual) connections. It would be useful to 
reformulate linearised gravity in terms of real connections and construct its
quantization. \\
\noindent (iii) Note that we have mainly been concerned with the kinematics
of $U(1)$ theory. The Fock representation, of course, supports the 
Maxwell Hamiltonian as an operator. 
Note that the normal ordering prescription adopted in \cite{lingrav}
is, of course, connected with Poincare invariance.
It is an open question as to how to 
express (presumably an approximation of) the Hamiltonian as an operator in the
Haar representation.\\
\noindent (iv) We have not been able to show continuity of the 
Fock PLF on ${\cal{HA}}_r$ or lack thereof. If the Fock PLF is continuous,
 a `Fock' measure,
$d\mu_F$, can be constructed on $\Delta_r$ and $\cal{H}$ can be identified 
with $L^2(\Delta_r, d\mu_F)$. 
The considerations of section 5b can then be extended to the $C^*$ algebras
${\overline{\cal{HA}}}_r$ and ${\overline{\cal{HA}}}$.

If, however, the Fock PLF on ${\cal{HA}}_r$
turns out {\em not} to be continuous, then a corresponding
Fock measure on $\Delta_r$ does not exist and it is incorrect to identify 
$\Delta_r$ with the `quantum configuration space'. If this is indeed the
case, then the emphasis on {\em continuous} cyclic representations of 
${\overline{\cal{HA}}}$  in loop quantum gravity \cite{ai} would seem 
unduly restrictive.\\
\noindent (v) The representation of kinematic loop quantum gravity is the 
$SU(2)$ counterpart of the Haar representation for $U(1)$ theory. 
An important
question is 
how the Fock space-graviton description of linearised gravity 
arises out of loop quantum gravity.
It is possible that some insight into this issue may be obtained by 
considering the following (simpler) question in the context of $U(1)$
theory. Is there any way in which an approximate Fock structure can be 
obtained from the Haar representation of $U(1)$ theory?
Since the PLFs play a key role in determining the type of 
representation, this work suggests that to get an approximate Fock 
structure, it may be a good strategy to try to approximate (in some, yet 
unknown way) the Fock PLF by the Haar PLF.

\noindent {\bf Acknowledgements}: I am very grateful to Jose Zapata for
 useful discussions and encouragement.

\section*{Appendix}
\subsection*{A1}

{\em Lemma 1}: Given \\
\noindent (i)  $\gamma_i \in {\cal L}_{x_0}$, $i=1..n$, $n$ finite,
 (ii) $A_a({\vec x})\in {\cal A}$
 and  (iii) $\epsilon >0 $, 
there exists a connection $A^{\epsilon}_a({\vec x})\in {\cal A}$ such that 
\begin{equation}
|H_{\gamma_{i(r)}}(A^{\epsilon}) - H_{\gamma_{i}}(A)| < \epsilon
\label{eq:hhdiff}
\end{equation}
for $i=1..n$. \\

\noindent
{\em Proof}: 
For a single loop $\gamma$, from (\ref{eq:xk})
\begin{equation}
|X^a_{\gamma}({\vec k})| < C_{\gamma} \;\;\; 
C_{\gamma}:={3\over (2\pi )^{3\over 2}}L_{\gamma}
\label{eq:xkbound}
\end{equation}
where $L_{\gamma}$ is the length of the loop as measured by the flat metric.
Since $A_a({\vec x})$ is Schwartz, we have, for arbitrarily large $N>0$,
\begin{equation}
|A_a({\vec k})| < {C_N \over k^N} \;\; {\rm for \; some}\; C_N>0.
\label{eq:akbound}
\end{equation}
From (\ref{eq:xkbound}) and (\ref{eq:akbound})
\begin{equation}
\int_{k> \Lambda}d^3k |X^a_{\gamma}(-{\vec k})A_a({\vec k})|
< {C_{N,\gamma}\over \Lambda^{N-3}}, \;\;
C_{N,\gamma}= {4\pi C_{\gamma} C_N\over N-1}.
\end{equation}
Thus, given $\delta >0$, there exists $\Lambda ( \gamma, \delta )$ such that
\begin{equation}
\int_{k> \Lambda(\gamma ,\delta )}d^3k |X^a_{\gamma}(-{\vec k})A_a({\vec k})|
< \delta .
\label{eq:xakbound}
\end{equation}

Let $f(k)>0$ be a smooth function such that 
\begin{eqnarray}
f(k) &=& e^{k^2r^2\over 2} \;\;{\rm for}\;\; k< \Lambda(\gamma ,\delta )\nonumber \\
  & < & e^{k^2r^2\over 2}\;\;{\rm for}\;\; 
\Lambda(\gamma ,\delta )< k< 2\Lambda(\gamma ,\delta )\nonumber \\
  & = & 1\;\;\;{\rm for}\;\; k> 2\Lambda(\gamma ,\delta )
\end{eqnarray}

Define $A^{\delta}_{a(r)}({\vec x})$ through its Fourier transform,
\begin{equation}
A^{\delta}_{a(r)} ({\vec k}) := f(k) A_a({\vec k}).
\label{eq:ark}
\end{equation}
Note that $A^{\delta}_{a(r)}({\vec x}) \in {\cal A}$.

From (\ref{eq:xrk}), (\ref{eq:ark}), and (\ref{eq:xakbound}) it follows that
\begin{equation}
|\int_{k> \Lambda(\gamma ,\delta )}d^3k 
X^a_{\gamma_{(r)}}(-{\vec k})A^{\delta}_{a(r)}({\vec k})|
< \delta .
\label{eq:xarkbound}
\end{equation}
From (\ref{eq:xrk}) and (\ref{eq:ark})
\begin{equation}
\int_{k< \Lambda(\gamma ,\delta )}d^3k 
X^a_{\gamma_{(r)}}(-{\vec k})A^{\delta}_{a(r)}({\vec k})
=\int_{k< \Lambda(\gamma ,\delta )}d^3k 
X^a_{\gamma}(-{\vec k})A_{a}({\vec k}).
\label{eq:xakxark}
\end{equation}
Using (\ref{eq:xakxark})
\begin{equation}
|H_{\gamma_{i(r)}}(A^{\epsilon}) -H_{\gamma_{i}}(A)  | =
|\exp i{\big(} \int_{k> \Lambda(\gamma ,\delta )}d^3k 
X^a_{\gamma_{(r)}}(-{\vec k})A^{\delta}_{a(r)}({\vec k})
-X^a_{\gamma}(-{\vec k})A^{a}({\vec k}){\big)}.
\label{eq:hdiff}
\end{equation}
From (\ref{eq:xakbound}), (\ref{eq:xarkbound}) and (\ref{eq:hdiff}), for 
small enough $\delta >0$, it can be seen that 
\begin{equation}
|H_{\gamma_{i(r)}}(A^{\epsilon}) -H_{\gamma_{i}}(A)  | < 4\delta
\end{equation}

For the loops $\gamma_i , \;i=1..n$,
\begin{eqnarray}
{\bar {\Lambda}}(\delta )& := &  \max_i \Lambda (\delta , \gamma_i )
\nonumber \\
{\bar A}^{\delta}_{a(r)}({\vec k}) & := & {\bar f} A_a({\vec k})
\end{eqnarray}
with 
${\bar f}(k)>0$  a smooth function such that 
\begin{eqnarray}
{\bar f}(k) &=& e^{k^2r^2\over 2}, \;\;{\rm for}\;\;
 k< {\bar \Lambda}(\delta )\nonumber \\
  & < & e^{k^2r^2\over 2}\; \;\;{\rm for}\;\;
{\bar \Lambda}(\delta )< k< 2{\bar \Lambda}(\delta )\nonumber \\
  & = & 1\;\;\;{\rm for}\;\;  k> 2{\bar \Lambda}(\delta )
\end{eqnarray}
Then, given $\epsilon >0$, choose some $\delta \leq {\epsilon \over 4}$ 
and set 
\begin{equation}
A^{\epsilon}_a({\vec k}) := {\bar A}^{4\delta}_{a(r)}({\vec k}).
\end{equation}
Then (\ref{eq:hhdiff}) holds.

\noindent {\em Lemma 2}: Given \\
\noindent (i) strongly independent loops $\gamma_i$, $i=1..n$, $n$ finite,
 (ii) $g_i \in U(1)$, $i=1..n$
 and  \\
\noindent
(iii) $\epsilon >0 $, 
there exists a connection $A^{\epsilon}_a({\vec x})\in {\cal A}$ such that 
\begin{equation}
|H_{\gamma_{i(r)}}(A^{\epsilon}) - g_i| < \epsilon
\end{equation}
for $i=1..n$. \\

\noindent
{\em Proof}: From \cite{al}, $A_a \in {\cal A}$  exists such that 
$H_{\gamma_{i}}(A) = g_i$, $i=1..n$. Therefore it suffices to construct
$A^{\epsilon}_a$ such that 
$|H_{\gamma_{i(r)}}(A^{\epsilon}) -H_{\gamma_{i}}(A)  | < \epsilon$.
But this is exactly the content of Lemma 1.

\noindent {\em Lemma 3}:Given
$A_a({\vec x})\in {\cal A}$, there exists 
$A_{a(r)}({\vec x})\in {\cal A}$ such that 
\begin{equation}
H_{\gamma_{(r)}}(A) =H_{\gamma}(A_{(r)}).
\end{equation}
for every $\gamma\in {\cal L}_{x_0}$.\\
\noindent {\em Proof}: From (\ref{eq:xrk}) and (\ref{eq:hr})
it immediately follows that the required 
$A_{a(r)}({\vec x})$ is determined by its Fourier transform via
$A_{a(r)}({\vec k})= e^{-k^2r^2\over 2}A_a({\vec k})$.

\subsection*{A2}

\noindent {\em Proposition}: The states 
$|\psi_{\{{\vec x}_i, {\vec v}_i\}}> :=
\prod_{i=1}^p{\hat O}_{{\vec x}_i, {\vec v}_i}|0>,\; (p=1,2..)$ together 
with 
$|0>$, span ${\cal D}_0$.

\noindent {\em Heuristic Proof}: The argument below is  a bit 
formal, but we expect
that it can be converted to a rigorous proof.

Define
\begin{equation}
|\psi, p> :=\int d^3k_1..d^3k_p \psi^{a_1..a_p}({\vec k}_1..{\vec k}_p)
(\prod_{i=1}^p
{\hat a}_{a_i}({\vec k}_i)+{\hat a}^{\dagger}_{a_i}(-{\vec k}_i))
|0>, 
\end{equation}
where $\psi^{a_1..a_p}({\vec k}_1..{\vec k}_p)$ has the same properties 
(a)-(c) (see section 4) as $\phi^{a_1..a_p}({\vec k}_1..{\vec k}_p)$.
$|\psi, p>$ 
along with $|0>$ span ${\cal D}_0$.
$|\psi, p>$ can be generated from $|\psi_{\{{\vec x}_i, {\vec n}_i\}}>$
as follows.

Note that from (\ref{eq:defo}),
\begin{equation}
{1 \over 2\pi^{3\over 2}}
\int d^3xe^{-i{\vec k}\cdot {\vec x}} {\hat O}_{{\vec x}, {\vec n}}=
{i({\vec n} \times {\vec k})^a   \over {\sqrt{2}} k} 
e^{-k^2r^2\over 2}
({\hat a}_a({\vec k})
         +{\hat a}^{\dagger}_a(-{\vec k})).
\label{eq:defoft}
\end{equation}
Define 
\begin{equation}
g^{a_1..a_p}({\vec k}_1..{\vec k}_p)
=(\prod_{i=1}^p e^{k_i^2r^2}{\sqrt{2}} k_i) 
\psi^{a_1..a_p}({\vec k}_1..{\vec k}_p).
\label{eq:defg}
\end{equation}
Given ${\vec k}_i\; i=1..p$, it is possible to construct a triplet
of vectors ${\vec u}^{i}_{a_i}({\vec k}_i)$ ($a_i= 1,2,3$ for each $i$),
such that
\footnote{ An explicit choice is as follows. Fix Cartesian coordinates 
$(x,y,z)$ and the corresponding unit vectors $(\hat{x},\hat{y},\hat{z})$. 
Then for $i=1..p$, 
${\vec u}^i_1 = {{\vec k}_i\times {\hat x}\over k}$,
${\vec u}^i_2 = {{\vec k}_i\times {\hat y}\over k}$,
${\vec u}^i_3 = {{\vec k}_i\times {\hat z}\over k}$.}
\begin{equation}
{({\vec u}^{i}_{b_i} \times {\vec k}_i)^{a_i}   \over k} =
\delta^{a_i}_{b_i} - {k^{a_i} k_{b_i} \over k^2} .
\label{eq:transverse}
\end{equation}
Then from 
(\ref{eq:defoft}), (\ref{eq:defg}) and (\ref{eq:transverse}), 
\begin{equation}
|\psi, p>=
\int (\prod_{l=1}^pd^3k_l
     \prod_{m=1}^pd^3x_m)\;
g^{a_1..a_p}({\vec k}_1..{\vec k}_p)
(\prod_{i=1}^p{e^{-i{\vec k}_i\cdot {\vec x}_i}\over 2\pi^{3\over 2}})
|\psi_{\{{\vec x}_i, {\vec u}^{i}_{a_i}\}}>
\label{eq:formal}
\end{equation}
It is in this formal sense that states of the type 
$|\psi_{\{{\vec x}_i, {\vec v}_i\}}>$ together with $|0>$ span 
${\cal D}_0$.

\noindent {\em Lemma 2}: 
${\hat O}_{{\vec x}, {\vec n}}$ is a densely defined, symmetric 
operator on the dense domain ${\cal D}_0$, which admits self 
adjoint extensions. 

\noindent {\em Proof}:
It is straightforward to check that 
\begin{eqnarray}
{\hat O}_{({\vec x}, {\vec n})}|\phi, p>
&=& 
\int (\prod_{i=1}^{p+1}d^3k_i) f^{a_{p+1}}(-{\vec k}_{p+1})
 \phi^{a_1..a_p}({\vec k}_1..{\vec k}_p)
(\prod_{j=1}^{p+1}{\hat a}^{\dagger}_{a_j}({\vec k}_j))
|0> \nonumber \\
&+& p\int (\prod_{i=1}^{p}d^3k_i) f_{a_{1}}({\vec k}_{1})
 \phi^{a_1..a_p}({\vec k}_1..{\vec k}_p)
(\prod_{j=2}^{p+1}{\hat a}^{\dagger}_{a_j}({\vec k}_j))
|0>
\label{eq:ophip}
\end{eqnarray}
where
\begin{equation}
f^a({\vec k}):= {i\over 2\pi^{3\over 2}}
e^{i{\vec k}\cdot {\vec x}} 
{({\vec n} \times {\vec k})^a   \over {\sqrt{2}} k} 
e^{-k^2r^2\over 2}.
\end{equation}
The  ultraviolet behaviour of $\phi^{a_1..a_p}({\vec k}_1..{\vec k}_p)$,
$f^a({\vec k})$ ensures that $||{\hat O}_{({\vec x}, {\vec n})}|\phi, p>||$
is finite. Thus  ${\hat O}_{({\vec x}, {\vec n})}$ is densely defined 
on ${\cal D}_0$.
By inspection ${\hat O}_{({\vec x}, {\vec n})}$ is also symmetric on 
${\cal D}_0$.  

To show existence of its self adjoint extensions, it is sufficient to
exhibit an antilinear operator $\hat C$ on $\cal F$ with 
${\hat C}^2= {\bf 1}$
which leaves ${\cal D}_0$ invariant and commutes with 
${\hat O}_{({\vec x}, {\vec n})}$ \cite{rsvon, amp}. As in 
\cite{amp}, take $\hat C$ to be the complex conjugation operator (in the 
standard Schrodinger representation) on  
${\cal F} = L^2 ({\cal S}^{\prime}, d\mu)$
where ${\cal S}^{\prime}$ is the appropriate space of tempered distributions
and $d\mu$ is the standard Gaussian measure, for free Maxwell theory. 
It can be seen that 
${\hat C}a_a({\vec k})= a_a(-{\vec k}){\hat C}$ and 
${\hat C}a^{\dagger}_a({\vec k})= a^{\dagger}_a(-{\vec k}){\hat C}$, and hence
${\hat O}_{({\vec x}, {\vec n})}{\hat C}={\hat C}
{\hat O}_{({\vec x}, {\vec n})}$.

\noindent {\em Lemma 3}:
\begin{equation} 
||{(e^{i{\hat D}} -1)\over i\pi\epsilon_m^2} |\psi >||\rightarrow 0.
\label{eq:dto0}
\end{equation}
as $\epsilon_m \rightarrow 0$. \\ 
\noindent {\em Proof:}
\begin{equation}
{i{\hat D}} = {i\int d^3k h^a({\vec k})}
               ({\hat a}_a({\vec k})
         +{\hat a}^{\dagger}_a(-{\vec k})) 
\label{eq:defeid}
\end{equation}
with 
\begin{equation}
h^a({\vec k})
= {e^{i{\vec k}\cdot {\vec x}}e^{-k^2r^2\over 2}\over 
                                         {\sqrt{2}}k 2\pi^{3\over 2}}
\big( \oint ds e^{i{\vec k}\cdot 
({\vec \gamma}^{\{m, {\vec x}, {\vec n}\}}-{\vec x})}
{\dot\gamma}^{a\{m, {\vec x}, {\vec n}\}}
-i \pi \epsilon_m^2 ({\vec n} \times {\vec k})^a
\big)
\end{equation}
A straightforward calculation, using \cite{gr},  shows that 
\begin{equation}
h^a({\vec k}) = 
{ie^{i{\vec k}\cdot {\vec x}}e^{-k^2r^2\over 2}\over 
                                      {\sqrt{2}}k2\pi^{3\over 2}}
({\vec n} \times {\vec k})^a\pi \epsilon_m^2
\big( 
{2J_1(\alpha_k \epsilon_m)\over \alpha_k \epsilon_m} -1
\big)
\label{eq:hj1}
\end{equation}
with $\alpha_k := |{\vec n} \times {\vec k}|$. 

Now, from (\ref{eq:defeid})
and (\ref{eq:comma}), 
\begin{equation}
e^{i{\hat D}}= e^{-\int d^3k |h^a({\vec k})h_a({\vec k})|}
               e^{i\int d^3k h^a({\vec k}){\hat a}^{\dagger}_a({\vec k})}
               e^{i\int d^3k h^a({\vec k}){\hat a}_a({\vec k})}.
\end{equation}
\begin{eqnarray}
\Rightarrow <\phi, p|e^{i{\hat D}}|\phi, p>
& =& e^{-\int d^3k |h^a({\vec k})h_a({\vec k})|}\int 
(\prod_{i=1}^pd^3k_i\prod_{j=1}^pd^3l_j)
\phi^{a_1..a_p}({\vec k}_1..{\vec k}_p)
\phi^{*b_1..b_p}({\vec l}_1..{\vec l}_p)  \nonumber \\
& &
<0|(\prod_{j=1}^p{\hat a}_{b_j}({\vec l}_j)-ih^*_{b_j}({\vec l}_j))
   (\prod_{i=1}^p{\hat a}^{\dagger}_{a_i}({\vec k}_i)+ih_{a_i}({\vec k}_i)) 
 |0>
\end{eqnarray}
\begin{eqnarray}
\Rightarrow <\phi, p|e^{i{\hat D}}-1|\phi, p> &=&  
          n!(e^{-\int d^3k |h^a({\vec k})h_a({\vec k})|}-1)
            \int \prod_{i=1}^pd^3k_i 
            | \phi^{a_1..a_p}({\vec k}_1..{\vec k}_p)|^2     
\nonumber \\
+ \;n!n e^{-\int d^3k |h^a({\vec k})h_a({\vec k})|}&&
\int (\prod_{i=2}^p d^3k_i) d^3k d^3l h_{a_1}({\vec k}) h^{*b_1}({\vec l})
          \phi^{a_1a_2..a_p}({\vec k},{\vec k}_2..,{\vec k}_p)
\phi_{*b_1a_2..a_p}({\vec l},{\vec k}_2..,{\vec k}_p)  
\nonumber\\
&+& O(h^4).
\label{eq:2terms}
\end{eqnarray}
Since 
$\big( 
{2J_1(\alpha_k \epsilon_m)\over \alpha_k \epsilon_m} -1
\big)$ is a bounded function, (\ref{eq:hj1}) implies that 
the $O(h^4)$ terms do not contribute to  (\ref{eq:dto0}) in the 
$\epsilon_m\rightarrow 0$ limit.

From Lemma 4 below and (\ref{eq:hj1})
the first term of (\ref{eq:2terms}) is of order $\epsilon_m^{5{1\over 2}}$
and the second is of order $\epsilon_m^5$. From this is it is clear that 
$||{(e^{i{\hat D}} -1)\over i\pi\epsilon_m^2} |\psi >||\rightarrow 0$
as  $\epsilon_m \rightarrow 0$.

\noindent {\em Lemma 4}:
Let $n$ be a positive integer and $g({\vec k})$ be a bounded function 
of rapid decrease (i.e. it falls to zero as $k\rightarrow \infty$,
faster than any inverse power of $k$.). Then, as $\epsilon\rightarrow 0$,
\begin{equation}
I:= |\int d^3k g({\vec k})
     \big( 
{2J_1(\alpha_k \epsilon)\over \alpha_k \epsilon} -1
\big)^n| < \; C\epsilon^{n-{1\over 2}}
\end{equation}
for some positive constant $C$ which depends on $n$ and $g$.\\
\noindent {\em Proof}:
\begin{equation}
I \leq 
\int_{k\leq\epsilon^{-{1\over 2}}} d^3k| g({\vec k})
     \big( 
{2J_1(\alpha_k \epsilon)\over \alpha_k \epsilon} -1
\big)^n| 
+
\int_{k >\epsilon^{-{1\over 2}}} d^3k| g({\vec k})
     \big( 
{2J_1(\alpha_k \epsilon)\over \alpha_k \epsilon} -1
\big)^n| .
\label{eq:j12}
\end{equation}
In the first term the range of integration is such that 
$\alpha_k \epsilon < \epsilon^{1\over 2}$. A straightforward 
calculation shows that the small argument expansion of 
$J_1(\alpha_k\epsilon )$ coupled with the rapid fall off property
of $g({\vec k})$ gives the bound
\begin{equation}
\int_{k\leq\epsilon^{-{1\over 2}}} d^3k| g({\vec k})
     \big( 
{2J_1(\alpha_k \epsilon)\over \alpha_k \epsilon} -1
\big)^n| \leq C_1(g,n) \epsilon^{n-{1\over 2}}.
\label{eq:c1gn}
\end{equation}
where $C_1(g,n)$ is a positive constant dependent on both $n$ and the 
properties of $g$. 

The rapid decrease property of $g({\vec k})$ ensures that, for small
enough $\epsilon$, the second term of (\ref{eq:j12}) falls off much faster 
than the first term. Hence, 
$I< \; C\epsilon^{n-{1\over 2}}$ where we have set $C:=2C_1(g,n)$.

\end{document}